%% file: sample-sigconf.tex
\documentclass[sigchi]{acmart}

\AtBeginDocument{%
  \providecommand\BibTeX{{%
    \normalfont B\kern-0.5em{\scshape i\kern-0.25em b}\kern-0.8em\TeX}}}

\setcopyright{acmlicensed}

\copyrightyear{2019} 
\acmYear{2019} 
\acmConference[AM'19]{Audio Mostly}{September 18--20, 2019}{Nottingham, United Kingdom}
\acmBooktitle{Audio Mostly (AM'19), September 18--20, 2019, Nottingham, United Kingdom}
\acmPrice{15.00}
\acmDOI{10.1145/3356590.3356596}
\acmISBN{978-1-4503-7297-8/19/09}

\usepackage{booktabs}
\usepackage{multirow}
\usepackage{amsmath}
\usepackage[]{algorithm2e}

\begin{document}

\title[A Real-Time Tempo and Meter Tracking System]{A Real-Time Tempo and Meter Tracking System \\for Rhythmic Improvisation}

\author{Filippo Carnovalini}
\email{filippo.carnovalini@dei.unipd.it}
\orcid{0000-0002-2996-9486}
\affiliation{%
  \institution{University of Padova (IT)}
}

\author{Antonio Rod\`a}
\email{roda@dei.unipd.it}
\affiliation{%
  \institution{University of Padova (IT)}
}


\begin{abstract}
Music is a form of expression that often requires interaction between players. If one wishes to interact in such a musical way with a computer, it is necessary for the machine to be able to interpret the input given by the human to find its musical meaning. In this work, we propose a system capable of detecting basic rhythmic features that can allow an application to synchronize its output with the rhythm given by the user, without having any prior agreement or requirement on the possible input. The system is described in detail and an evaluation is given through simulation using quantitative metrics. The evaluation shows that the system can detect tempo and meter consistently under certain settings, and could be a solid base for further developments leading to a system robust to rhythmically changing inputs.
\end{abstract}

\ccsdesc[500]{Human-centered computing~Interactive systems and tools}
\ccsdesc[500]{Applied computing~Sound and music computing}
\ccsdesc[300]{Information systems~Music retrieval}

\keywords{rhythm, improvisation, musical interaction, music information retrieval}


\maketitle

\input{content.tex}

\vspace{-2mm}
\begin{acks}
F.C. is funded by a doctoral grant by University of Padua.
\end{acks}
\vspace{-2mm}
\bibliographystyle{ACM-Reference-Format}
\bibliography{am19}



\end{document}

%% file: content.tex
\section{Introduction}

Music is a form of expression that favors the interaction between musicians: in order to perform a non-solo piece, they need to listen and adapt to each other, especially on a rhythmic level.
If one wants to enable a similar interaction with a machine, it is necessary to give it the capability of listening to a musical input and to interpret it in a musical way, as a human would do. For example, it would be desirable that software was able to follow a human improvisation. In this direction, there is research on score following and automatic accompaniment (among many works, see e.g.~\cite{muller_towards_2004,Cristani:2010aa})
, but these hardly account for deviations from a pre-defined standard, and cannot adapt to musical improvisation at all. 
What we try to do in the work presented in this paper is to allow the computer to analyze in real-time rhythmic features of input, for instance, received via MIDI protocol, making the computer able to understand the rhythmic sense of what is being played.

This could be applied to automatic music generation: a computer system could generate music in real-time following the rhythm established by the human user, or it could simply play back a pre-recorded piece that is selected from a database to fit the input rhythm. Another possible application is to train musicians to perform certain rhythmic patterns: if the system is able to correctly detect the performed rhythm it is also possible to give feedback on how to correct rhythmic errors. 
As a final example of application, the proposed system has already been applied to a musical serious game inspired by music therapy, where two players must collaboratively create a rhythmic pattern, by means of two MIDI pads or miked drums~\cite{dafx10b}, in order to score points and receive a more pleasing musical augmentation to their rhythmic performance~\cite{carnovalini}. 

\subsection{Related Works}

The basic requirements of the system we will describe are: (1) being able to infer the tempo and meter (i.e., the underlying time signature) played by the user, (2) being able to perform in real-time actions that are synchronized with the rhythm of the user. 
The first requirement is something that can be easily found in the literature under the field of Music Information Retrieval: without the requirement of real-time analysis, these are basic information that are normally of interest for the musical analysis of a digital score or an audio file~\cite{dixon_automatic_2001,frieler_beat_2004,whiteley_bayesian_2006,gouyon_determination_2003}. The second requirement is instead more frequently considered by research relating to score-following and automatic accompaniment~\cite{muller_towards_2004,raphael_bayesian_2002}.

In particular, those works that are focused on following a human improvisation best cover our requirements, but often make assumptions on the input performance, requiring some prior knowledge before the improvisation starts. One example is the MIDI Accompanist by Toiviainen~\cite{toiviainen_interactive_1998}, that uses oscillators to adapt to a non-predefined beat. This work has a complex mathematical basis, and it only focuses on the adaptation to a beat, not considering meter. Another approach is that of Beatback~\cite{hawryshkewich_beatback:_2010}, which is meant to follow drummers in their improvisation, but the synchronization of the system is based on the fact that the tempo is decided before the improvisation, thus making it not adaptive. A similar approach is that of B-keeper~\cite{robertson_b-keeper:_2007}, that analyzes the audio of a kick drum to perform synchronization, but again requires the prior knowledge of the approximate tempo and does not consider meter. The improvisation follower by Xia and Dannenberg~\cite{xia_improvised_2017} is based on melody instead, but is limited to improvisation over a given melody and the system needs to be trained over examples of the target accompaniment.  
Rather than starting from these works, that do not consider the detection of all the features we desire, we started from a Music Information Retrieval algorithm~\cite{frieler_beat_2004} that we adapted and expanded to work in real-time.\\

\vspace{-4mm}
\section{Theoretical Basis}
\label{sec:teo}

\begin{figure*}[t]
 \centering
 \includegraphics[trim={1.0cm 22.1cm 1.6cm 1.7cm}, clip, width=0.9\textwidth]{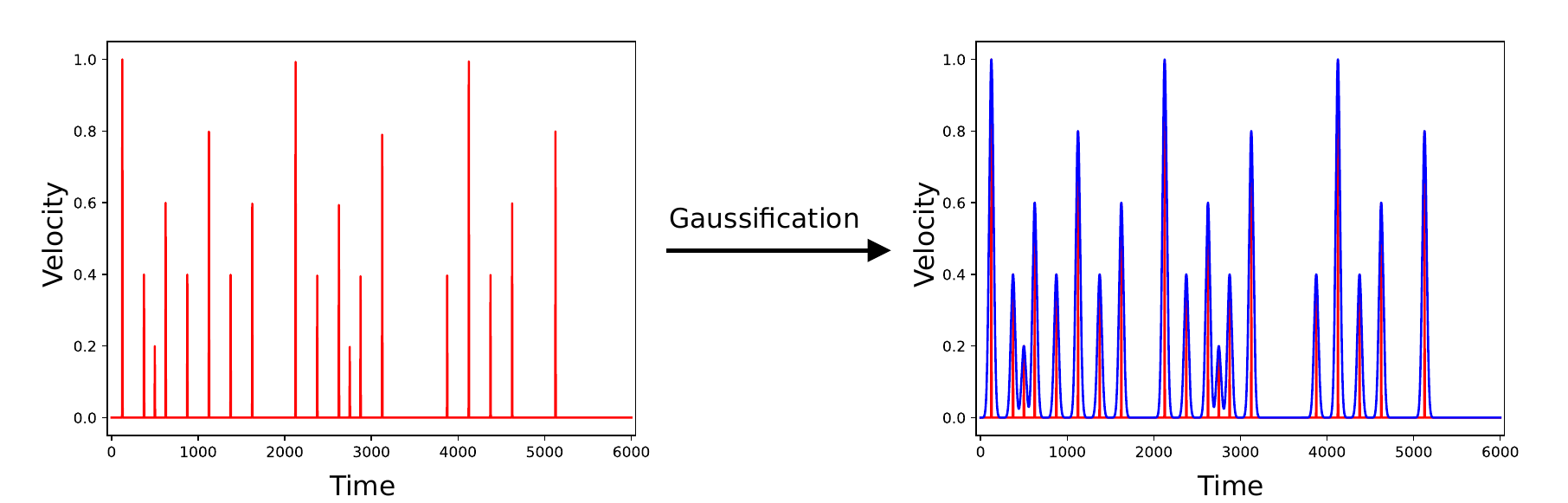}
 \caption{Example of Gaussification. On the left, a set of time points with velocities are represented as spikes over time. On the left, a Gaussification is applied to the same points, to obtain a integrable function.}
 \label{fig:gauss}
\end{figure*}
The proposed algorithm is based on the concept of Gaussification, as introduced by Frieler~\cite{frieler_beat_2004}. The algorithm's input is a list $R$ of $N$ time points (onsets of notes) and a list $V$ of $N$ coefficients (velocities of the same notes). These could be received through MIDI protocol, for example by playing on a velocity-sensitive MIDI pad, or could be inferred from an audio source extracting energy peaks. 
The system then constructs the linear combination of Gaussians centered at the time points in $R$, each multiplied by the corresponding coefficient in $V$ (as an example, see Figure \ref{fig:gauss}). 
To define Gaussification more precisely: 

Let $R = \{t_i\}_{1\leq i\leq N}$ and $V = \{v_i\}_{1\leq i\leq N}$, and $\sigma$ the standard deviation of the Gaussian kernel.\\ Then 
$$G_{R,V}(t) = \sum^{N}_{i=1}v_i e^{-\frac{(t-t_i)^2}{2\sigma ^2}}$$
is the Gaussification of $R,V$. Throughout the work, $\sigma$ was fixed to 25~ms, as suggested by Frieler~\cite{frieler_beat_2004}.

The idea behind this representation of the input is that human players do not follow precisely the beat as a metronome would do, but the timing of each human-played note is an approximation of that regular beat. The Gaussification allows the algorithm to consider this approximation when computing other features. 

The Gaussification has the advantage of being an integrable function. Thanks to that, Frieler~\cite{frieler_beat_2004} was able to define the correlation between two Gaussifications as follows: 
$$CG_{(R_1,V_1),(R_2,V_2)}(t) = \sum^{N_1}_{i=1}\sum^{N_2}_{j=1}v^1_i v^2_j e^{-\frac{(t-(t^1_i-t^2_j))^2}{2\sigma ^2}}$$
Similarly, one can compute the autocorrelation of a single Gaussification by computing its correlation with a time-shifted version of itself:
$$AG_{R,V}(t) = \sum^{N}_{i,j=1}v_i v_j e^{-\frac{(t-(t_i-t_j))^2}{2\sigma ^2}}$$

The main difference between the formulations that are reported here and the original ones by Frieler~\cite{frieler_beat_2004} is that the formula for the Gaussian used here is not the usual probability density function. This variant was chosen so that the peak of each Gaussian centered in $t_i$ is exactly $v_i$, giving a more precise representation of the velocity of each note.

Another important function needed for the meter and beat detection algorithm is one that allows choosing the most likely tempo when more than one is possible. If a song has a tempo of 100~bpm, it is possible to perceive it as 50~bpm or 200~bpm: if one makes such a mistake tapping along the song, they would correctly tap to beats of the song, but either skipping one every two beats or by tapping twice every beat. Since this does not result in a loss of synchronization this is not to be considered a major mistake, but it is nonetheless necessary to choose the most likely tempo in these ambiguous situations. Psychology of rhythm comes in handy for this task. In particular, Frieler~\cite{frieler_beat_2004} used a function derived from Parncutt's pulse-period salience~\cite{parncutt_perceptual_1994}:
\begin{equation} P(t) = e^{-2\log^2_2 t/t_s}
\label{eq:parncutt}\end{equation}
where $t_s$ is the "spontaneous tempo" expressed in milliseconds, that is the tempo that humans are more likely to tap to if instructed to regularly tap to a tempo of their own choice. Throughout the research, $t_s$ was fixed to 500~ms (120~bpm).

\section{Algorithm}
\label{sec:alg}

This section describes the procedures used to  estimate the tempo and meter, and also to predict the onset of measures needed to effectively follow the rhythm established by an improvising human.

\subsection{Beat Estimate}

Algorithm \ref{alg:beat} shows how the tempo is estimated. Of the input set of time points, only the ones played in the last 6000~ms are kept. 
This time window was chosen because this is large enough to include 2 measures in most tempo settings, and because a bigger window would increase computational time (see Section \ref{sec:realtime}), but other settings could be tested, for example by changing the window size dynamically.
The autocorrelation of the gaussification of the points within this time window is computed as described in the previous section. The rationale behind this is that a musical rhythm will be highly similar to itself when shifted by the duration of a beat, thus the autocorrelation of the rhythm will peak on the most reasonable beat candidates.
The possible beat durations considered are between 100~ms and 2000~ms, corresponding to tempos between 600~bpm and 30~bpm. For each possible tempo, the autocorrelation is normalized by dividing the result with the autocorrelation of the signal with itself, that is greater or equal to any other possible autocorrelation value. This gives a result in the [0,1] interval, that is then multiplied by the value of function \eqref{eq:parncutt} for the same tempo. This weighting makes the algorithm prefer tempos that are closer to the perceived preferred tempo. The amount of time that gives the maximum result is the chosen duration of a beat. To obtain a value in bpm one simply needs to use the following equation:
$$bpm = \frac{60000}{beat~duration}$$

The algorithm also computes the clarity, that is simply the value of the autocorrelation of the chosen tempo before the application of \eqref{eq:parncutt}. This is a value in the [0,1] interval and can be seen as the confidence of the algorithm in stating that the chosen beat is the correct one. 
\begin{algorithm}[ht]

\SetAlgoLined
\KwIn{$R, V, window, current\_time$}
$N \gets$ length($R$)\;
\For{$i \gets 0$ \KwTo $N$}{
 \eIf{$R[i]<$ $current\_time - window$}{
$R$.delete(i), $V$.delete(i)\;
 }{
$R[i] \gets$ $R[i]-$($current\_time - window$)\;
 }
}
$norm \gets AG_{R,V}(0)$\;
\For{$i \gets 100$ \KwTo $2000$}{
$A[i] \gets$ $(AG_{R,V}(i)/norm) * P(i)$\;
}
$beat \gets argmax(A)$\;
$clarity \gets max(A)/P[beat]$\;
\Return{$beat,clarity$}\;

 \caption{Estimation of the beat duration, along with its clarity.}\label{alg:beat}
\end{algorithm}

\subsection{Meter Estimate}

\begin{figure}[t]
 \centering
 \includegraphics[ width=0.4\textwidth]{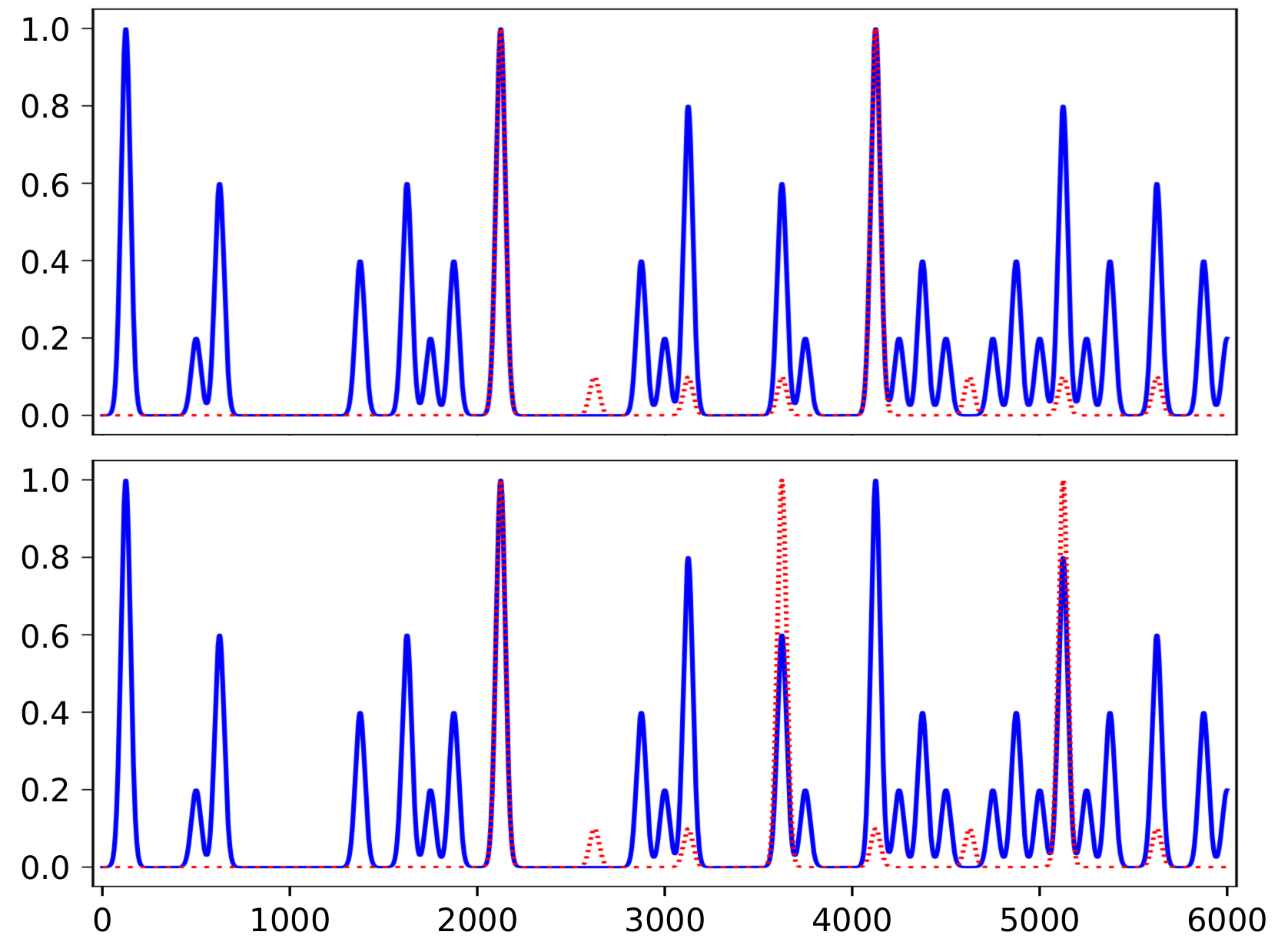}
 \caption{The gaussification of an input set of time points (solid blue line) is compared to the prototypes for 4/4 (top) and 3/4 (bottom) meters (red dotted lines). The prototype's phases computed by the system translated them to maximize the correlation with the input, but 4/4 was (correctly) chosen in this case as it scores a higher correlation.}
 \label{fig:proto}
 \vspace{-4mm}
\end{figure}

Once the beat is estimated, the meter is computed based on the assumption that meter emerges as a pattern of accents~\cite{povel_perception_1985}. 
A series of rhythmic patterns are predefined, and a "prototypical" signal is constructed for each by generating time points distanced by the estimated beat, having a velocity defined by the rhythmic pattern, following the procedure in Algorithm \ref{alg:proto}. 
After these prototypes are generated, the meter is estimated by choosing the prototype that has the highest correlation to the input time points set. 
In this implementation we only considered $\frac{3}{4}$ and $\frac{4}{4}$ as possible meters, but it would be easy to extend the procedure to other meters using additional prototypes. Exactly nine time points are generated so that the sum of the velocities of all the points are the same between the two prototypes, meaning that the correlations computed for the two meters are comparable.
Since there is no guarantee that the beginning of the window coincides with the beginning of a measure, the the correlation is computed giving a variable time shift as input, that ``moves'' the prototype along the input points.
The time shift that results in the highest correlation is the ``phase'' of the signal, i.e. the beginning of a measure in the window, as shown in Figure \ref{fig:proto}. The meter whose prototype has the highest correlation (considered the phase) is the chosen estimate for meter.
The complete procedure is reported in Algorithm \ref{alg:meter}, that uses the function $generatePrototype()$ described by Algorithm \ref{alg:proto}.

\begin{algorithm}[ht]

\SetAlgoLined
\KwIn{$meter, beat$}
$R_{meter} \gets []$, $V_{meter} \gets []$\;
\For{$i \gets 0$ \KwTo $8$}{
 $R_{meter}.append(i* beat)$\;
 \eIf{$modulo(i,meter)==0$}{
$V_{meter}.append(1)$\; 
 }{
$V_{meter}.append(0.1)$\;
 }
}
\Return{$R_{meter},V_{meter}$}\;
 \caption{Generation of a prototype.}\label{alg:proto}

\end{algorithm}

\begin{algorithm}[ht]

\SetAlgoLined
\KwIn{$beat, R, V, window, current\_time$}
$R_3,V_3 \gets generatePrototype(3,beat)$\;
$R_4,V_4 \gets generatePrototype(4,beat)$\;
\For{$i \gets 0$ \KwTo $window$}{
$corr_3[i] \gets$ $CG_{(R,V),(R_3,V_3)}(i)$\;
$corr_4[i] \gets CG_{(R,V),(R_4,V_4)}(i)$\;
}
\eIf{$max(corr_3)>max(corr_4)$}{
$meter \gets 3$\;
}{
$meter \gets 4$\;
}
$phase \gets argmax(corr_{meter})$\;
\Return{$meter,phase$}\;

 \caption{Estimation of the meter, along with the phase.}\label{alg:meter}
\end{algorithm}

\subsection{Prevision of Next Measure}

From the other procedures, the system already has an estimate of the duration of a beat and of the number of beats in a measure, thus can easily obtain the duration of a measure:$$measure = beat \cdot meter$$
In order to be synchronized with the human improvisation, the system needs a way to determine when to expect the beginning of a measure, just like a human joining an improvisation would wait for the beginning of a measure. 
It is possible to forecast the beginning of first measure outside the considered window with the following: 
$$onset = current\_time + measure- ((window-phase)\%measure)$$
Where \% is the modulo operation, and $window$ must be the same time window used in the above procedures. 
Ideally, the start of the next measure is obtained by adding the duration of a measure until the result is beyond the window. In practice, the same result is achieved implicitly via the modulo operation.


\section{Evaluation}
\label{sec:eval}

The evaluation of the system was carried out through a variety of simulations, trying to quantitatively assess if and how well the  requirements described in the Introduction are met by the system. 
The first experiment we describe relates to the last requirement: being able to function in real-time. 

\subsection{Experiment 1: Real-Time}
\label{sec:realtime}

Before describing the experiment in itself, it is useful to make a few considerations on the implementation of the system that are relevant to real-time computation.
All the procedures for the estimation of the rhythmic features that are described above allow an easy implementation as stateless functions. Having stateless functions allows an easy delegation of these computations to an asynchronous process different from the one that collects the input data in real-time. The $current\_time$ argument, passed to this process when the computation is required, is crucial for this: the code should never modify this value, so that the time passing during the computation does not change the result. When the result is available to the main process, it is easy to check if the estimated beginning of the next measure has already passed or not.
Practically, envisioning an application that needs to synchronize with an input rhythm, the rhythmic analysis process should be called at constant time intervals. For example, two or three times every second (every 500 or 333~ms). 
To consider the requirement satisfied, we must make sure that the computation is carried out before another call is made (i.e., it must last less than 333 or 500~ms). 

\begin{table}[t]
\begin{tabular}{@{}l|rrr@{}}
\toprule
Notes & \multicolumn{1}{c}{Avg. Time [s]} & \multicolumn{1}{c}{SD} & \multicolumn{1}{c}{\% SD} \\ \midrule
30 & 162.62 & 2.17 & 1.34  \\
35 & 215.32 & 3.15 & 1.46  \\
40 & 253.98 & 4.37 & 1.72  \\
45 & 315.04 & 5.30 & 1.68  \\
50 & 360.70 & 6.20 & 1.72  \\
55 & 432.86 & 11.37& 2.63  \\
60 & 486.32 & 11.03& 2.27  \\
65 & 568.56 & 3.44 & 0.61  \\
70 & 640.88 & 4.15 & 0.65  \\
75 & 705.24 & 12.18& 1.73  \\
80 & 788.92 & 15.30& 1.94  \\ \bottomrule
\end{tabular}
\caption{Average execution time (in seconds) for the extraction of tempo, meter and beginning of next measure, along with absolute and percent standard deviation (SD). The timings were measured on a 2013 Macbook Pro (2.4GHz Intel i5 processor, 4GB Ram)}
\label{tab:time}
\vspace{-10mm}
\end{table}

\subsubsection{Method}
The computational time of the algorithms described above grows as $O(N^2)$, where $N$ is the size of the input arrays of time points and velocities, or more simply put the number of input notes. This depends on the user improvisation and on the size of the window.
For this simulation, we fixed the size of the window to 6000~ms, and generated regular rhythms having from 30 to 80 notes (with a step of 5) inside the 6000~ms time window, and run the algorithm for the estimate of tempo, meter and onset of next measure 50 times per each setting on a 2013 Macbook Pro (2.4GHz Intel i5 processor, 4GB Ram). The results are shown in Table \ref{tab:time}.

\subsubsection{Discussion}
Despite the low computational power of the used machine, the algorithm is able to analyze a set of 45 notes in less than 333~ms, and up to 60 notes in less than 500~ms. To give an idea, if the user keep a tempo of 120~bpm beating every 16th note, in 6000~ms he would play 48 notes, that is fine if the system is called twice a second but is borderline when invoked three times per second. Faster tempos would require too much time, thus making the system fail. It could be a good idea to add a limit to the number of notes to be analyzed, beside the time limit imposed by the window.

\subsection{Experiment 2: Steady Tempo}
\label{sub:exp2}


The second experiment aims to compute metrics relating to the ability of the system to predict the tempo and meter, assuming a quasi-steady rhythmic production.  
The performance is evaluated using as input the rhythms generated by a simulator, able to create random event, characterized by reasonable (in the sense of human-like) constant tempo and meter, given as parameters.

\begin{figure}[t]
\includegraphics[trim={0cm 0.3cm 0cm 0.2cm}, clip,width=0.9\linewidth]{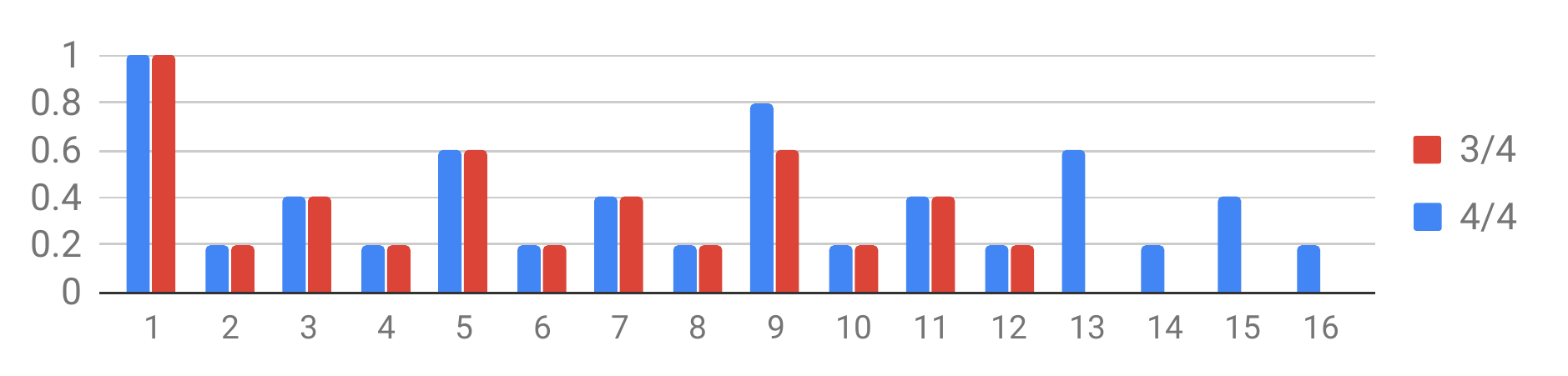}
\vspace{-3mm}
\caption{The importance assigned by the simulator to each of the 16th notes positions in a measure, both for 4/4 and 3/4 meters.}
\label{tab:gen}
\vspace{-4mm}
\end{figure}

\subsubsection{Simulations Setup}
 The simulation used does \emph{not} work in real-time, but it simulates the passing of time via an internal metronome implemented as an integer counter. Iteratively, the simulator consults, for every 16th note, a predefined table determined by the chosen meter, whose cells represent the importance of each metric position in the measure. The used weights are reported in Figure \ref{tab:gen}. 
Algorithm \ref{alg:simulator} shows how the table is used to generate a rhythm in $\frac{4}{4}$ meter (but is easy to adapt to $\frac{3}{4}$): for every 16th note the content of the table represents both the probability of generating a note corresponding to that time point and the velocity associated to that note if generated. Both Figure \ref{fig:gauss} and \ref{fig:proto} are examples of $\frac{4}{4}$ rhythms generated in this way.  
The algorithm includes the possibility of adding a random error to the generated time, to simulate the human imprecision in keeping a rhythm, that depends on the $\sigma_{err}$ parameter.

\begin{algorithm}

\SetAlgoLined
\KwIn{$beat$}
$metro \gets 0$\, $time \gets 0$\;
\For{$i \gets 0$ \KwTo $160$}{
 $importance \gets table[metro]$\;
 \If{$random(0,1)<importance$}{
$R.append(time + error(\sigma_{err}))$\;
$V.append(importance)$\;
 }
 $metro \gets modulo((metro+1),16)$\;
 $time \gets (time+beat)$\;
}
\Return{$R,V$}\;
 \caption{Generation of a simulated rhythm in 4/4.}\label{alg:simulator}
\end{algorithm}

\vspace{-4mm}
\subsubsection{Metrics}

Every time a note is generated, an estimate for tempo, meter, and onset of next measure is done and compared to the correct value, to compute the following metrics:
\begin{description}
\item[Tempo Accuracy (T-AC)]: a score of 100 is assigned to each correct tempo estimate, a score of 75 is instead given if the estimate is half or double the real tempo, as this is considered a minor mistake. A tolerance of 10~ms is considered in both cases. A score of 0 is given otherwise. The final value is the average score;
\item[Meter Accuracy (M-AC)]: the percent of times the estimated meter coincides with the real meter;
\item[Precision (P)]: computed as the ratio of estimates that are within 50~ms from an actual onset of a measure over the total amount of estimates;
\item[Recall (R)]: computed as the ratio of actual onsets of measures that are estimated within 50~ms over the total number of measures minus one (as the very first measure is impossible to forecast). 
\end{description}

Precision and Recall are computed only when the simulation is over, while Tempo and Meter Accuracy are computed progressively during the simulation. This is important because the ``real'' tempo and meter are the ones that the simulator is considering at the moment of the estimate, and it could change over the simulation for some experiments.  

\subsubsection{Experiment Setup}
For this experiment the simulator module was instantiated varying two parameters. The first was the tempo, that varied from 60~bpm to 200~bpm with 20~bpm intervals. The second parameter is $\sigma_{err}$: when generating the onsets of the rhythmic events, a random error based on the Normal distribution was added. $\sigma_{err}$ is the standard deviation of the distribution, that imitates different rhythmic precision levels of the simulated human player~\cite{repp_sensorimotor_2005}.
For each setting, the simulator ran 50 times, each on a freshly-generated random rhythm, and the metrics were averaged over these executions. The tempo did not change during an execution, and the meter was fixed to $\frac{4}{4}$ for all the experiment. 

The goal of this experiment was to test how well the algorithm responded to rhythmic imprecision, and if the chosen tempo was relevant to the effectiveness of the algorithm. The results of these experiments are reported in Tables \ref{tab:variances} and \ref{tab:tempos}.

\begin{table}[t]
\begin{tabular}{@{}l|rrrr@{}}
\toprule
\multicolumn{1}{c|}{$\sigma_{err}$ [ms]} & \multicolumn{1}{c}{T-AC (SD)} &  \multicolumn{1}{c}{M-AC (SD)}  & \multicolumn{1}{c}{P (SD)} &  \multicolumn{1}{c}{R (SD)}  \\ \midrule
0 & 79.6 (19.1) &87.8 (10.3)& 54.8 (19.6)& 86.4 (14.5)\\
2.5&79.5 (17.2)& 87.4 (10.5)& 53.3 (19.0)& 85.8 (13.9)\\
5 & 76.0 (18.6)& 87.6 (11.5)& 49.9 (20.0)& 83.1 (15.9)\\
7.5&70.5 (20.4)& 86.1 (11.5)& 45.6 (19.7)& 80.1 (15.9)\\
10& 66.7 (21.2)& 86.0 (12.2)& 42.5 (19.3)& 77.0 (17.7)\\
15& 53.9 (23.3)& 84.8 (11.7)& 34.5 (17.5)& 70.3 (18.8)\\
20& 41.2 (22.5)& 83.0 (11.8)& 25.8 (15.3)& 59.7 (19.8)\\
25& 33.4 (20.5)& 81.1 (12.1)& 21.2 (14.3)& 53.0 (20.4)\\ \bottomrule
\end{tabular}
\caption{The four metrics (with standard deviation) computed over the various trials for each of the $\sigma_{err}$ settings, averaged across all tempo settings.}
\label{tab:variances}
\vspace{-9mm}
\end{table}

\begin{table}[t]
\begin{tabular}{@{}l|rrrr@{}}
\toprule
Beat [ms]([bpm])  & \multicolumn{1}{c}{T-AC (SD)} & \multicolumn{1}{c}{M-AC (SD)} & \multicolumn{1}{c}{P (SD)}& \multicolumn{1}{c}{R (SD)} \\ \midrule
1000 (60)       & 56.9 (18.2)  & 93.5 (5.6) & 30.1 (12.5)  & 76.8 (18.3)  \\
750 (80)        & 58.0 (24.6)  & 88.4 (7.2) & 32.5 (15.0)  & 72.8 (19.1)  \\
600 (100)       & 78.1 (22.6)  & 83.2 (6.9) & 45.5 (17.0)  & 83.5 (17.6)  \\
500 (120)       & 81.9 (19.9)  & 94.7 (5.8) & 66.2 (21.0)  & 88.5 (13.7)  \\
428 ($\sim$140) & 78.1 (23.0)  & 92.9 (7.1) & 52.0 (20.5)  & 86.1 (16.7)  \\
375 (160)       & 45.8 (28.5)  & 83.4 (13.2)& 19.3 (9.6)   & 49.8 (16.1)  \\
333 ($\sim$180) & 44.3 (21.8)  & 74.0 (9.5) & 34.5 (15.8)  & 66.7 (18.2)  \\
300 (200)       & 57.6 (18.99  & 73.8 (10.9)& 47.3 (19.1)  & 71.0 (17.3)  \\ \bottomrule
\end{tabular}
\caption{The four metrics (with standard deviation) computed over the various trials for each of the tempo settings, averaged across all $\sigma_{err}$ settings.}
\label{tab:tempos}
\vspace{-7mm}
\end{table}

\subsubsection{Discussion}
In the best-performing conditions, i.e. either having very low $\sigma_{err}$ or having a tempo of 120~bpm, the system performs really well: both the tempo and meter accuracy are near or above 80. The recall in estimating the beginning of new measures is generally much higher than the precision, and is generally lower than what one could expect from the value of the tempo and meter accuracy: if the tempo and meter are correctly identified it should not be difficult to foresee a new measure. This difference can be explained in two ways. 
First, the measure onset determination procedure may not be precise enough, meaning that sometimes the beginning of a measure is estimated to be one of the weaker beats and not the first one. Second, the fact that a double tempo is considered as nearly correct means that the prevision might fall on half a measure rather than a full one. By looking more qualitatively at some of the generated rhythms where there are errors, it seems that the sum of these two factors makes it happen that the prevision often falls on the third and sometimes on the second or fourth beat.

When deviating from the best settings, the performances decrease, especially concerning the tempo accuracy and the precision. 
Changing the tempo from the ideal one, the performance of the tempo accuracy degrades, probably because of the Parncutt function~\eqref{eq:parncutt} the double or half tempo estimates become more frequent as the tempo becomes more extreme. Yet, the change is different for slow and fast tempos: faster tempos perform worse. This is probably due to the fact that the error on the input given by $\sigma_{err}$ becomes more evident as the intervals between notes become shorter. 

As $\sigma_{err}$ grows, the performances worsen (although the meter accuracy remains over 80), but this comes to little surprise. With $\sigma_{err}\leq$15~ms (as expected from a musically trained human, although this depends on many factors~\cite{repp_sensorimotor_2005}), the tempo estimate remains correct more than half of the times, but having $\sigma_{err}>$20~ms (which is expected from a non-musician playing a steady rhythm) is below that threshold. This, coupled with the fact that the precision is rather low, means that additional controls might be needed to keep the estimate and the ability to follow the improvisation consistent, for example taking the median of a series of consecutive estimates.

\subsection{Experiment 3: Sudden Changes}

The above experiment only evaluated the general effectiveness of the proposed method when a fixed tempo and meter is kept throughout the execution. This experiment tries to check how quickly the system can adapt if those values suddenly change.

\subsubsection{Experimental Setup}
This experiment uses two simulations: one in which the meter changes, and one where the tempo changes. In both scenarios, the simulation performs five measures without any change, and then suddenly changes setting. From that moment, the time that passes until the system has predicted the new tempo or meter correctly ten times is measured, meaning that the system has successfully adapted. 

When changing the meter, the simulation was run with three tempo settings to see how the estimate was affected, both going from a $\frac{4}{4}$ to a $\frac{3}{4}$ meter and vice-versa. The results, averaged over 50 runs for each setting, are shown on Table \ref{tab:metro_change}.
When changing the tempo instead, the initial tempo was fixed to 500~ms (120~bpm), but the amount of change varied from 50 to 200~ms (both increasing and decreasing the speed). Each setting was tested on 50 runs. The results are reported on Table \ref{tab:tempo_change}.
The $\sigma_{err}$ parameter was kept to 10~ms.
Since the reported timings are highly dependent on the tempo, the results are also reported as number of measures.

\begin{table}[t]
\begin{tabular}{llll}
\toprule
 & \multicolumn{1}{c}{Tempo [ms]} & \multicolumn{1}{c}{Avg. Time [ms] (SD)} & Measures \\ \hline
\multirow{3}{*}{$\frac{4}{4} \rightarrow \frac{3}{4} $} & 375& 6598 (2355)  & 5.87 \\
 & 500& 4625 (1488)  & 3.08 \\
 & 750& 5662 (2013)  & 2.52 \\ \hline
\multirow{3}{*}{$\frac{3}{4} \rightarrow \frac{4}{4} $} & 375& 6257 (2376)  & 4.17 \\
 & 500& 5047 (976)& 2.52 \\
 & 750& 3465 (924)& 1.16 \\ \bottomrule
\end{tabular}
\caption{The time needed by the algorithm to detect a change of meter, in milliseconds and in number of measures.}
\label{tab:metro_change}
\vspace{-8mm}
\end{table}

\begin{table}[t]
\begin{tabular}{@{}rll@{}}
\toprule
\multicolumn{1}{l}{$\Delta$ Beat [ms]} & \multicolumn{1}{c}{Avg. Time [ms] (SD)} & \multicolumn{1}{c}{Measures} \\ \midrule
-50 & 7463 (1158)  & 4.15 \\
50  & 7711 (1775)  & 3.51 \\
-100& 6557 (1778)  & 4.10 \\
100 & 7995 (1918)  & 3.33 \\
-150& 7136 (1315)  & 5.10 \\
150 & 8866 (2347)  & 3.41 \\
-200& 5537 (1213)  & 4.61 \\
200 & 9971 (3048)  & 3.56 \\ \bottomrule
\end{tabular}
\caption{The time needed by the algorithm to detect a change of tempo (millisecond added or subtracted from each beat), in milliseconds and in number of measures.}
\label{tab:tempo_change}
\vspace{-10mm}
\end{table}

\subsubsection{Discussion}
As evident from the results, sudden changes do not lead to immediate adapting. Even for a human it is not easy to immediately detect a change in meter before listening to at least one full measure to realize that the metric accents have changed. That considered, the results for meter changes are satisfactory. 

For a human, noticing that the tempo has changed is usually more immediate, but instead our system performs worse on this task, especially when the tempo becomes faster (denoted on the tables by tempo intervals with the minus sign, as the bpm value is inversely related to the beat duration). 
The difficulty of the adaptation to tempo changes is probably due to the fact that the time window of considered notes is fixed to 6000~ms, and thus will consider the old tempo until the notes relating to that tempo are outside the window. Windows that vary in size could possibly be helpful to this task. 

\subsection{Experiment 4: Gradual Tempo Change}

The final experiment tests how the system reacts when the tempo grows not in a sudden way but gradually over time. 

\subsubsection{Experimental Setup}
The setup for this experiment was similar to that of Experiment 2, but here instead of having a fixed tempo, there were five measures with a fixed tempo of 500~ms, then a measure where the tempo incremented or decremented by a fixed amount every 16th note, followed by four measures where the reached tempo remained unchanged (the final tempo is $500~ms\pm increment \cdot 16$). The possible values for the tempo change were 1 to 5~ms, and for each 50 runs were made increasing the tempo and 50 runs were made decreasing the tempo. The $\sigma_{err}$ parameter was kept to 10~ms.
The averaged results, along with a baseline where the tempo does not change, are reported in Table \ref{tab:increment}. The results are averaged over over 50 runs for each setting. There was no significant difference between the increasing tempo versus the decreasing one, so the two cases were joined.

\begin{table}[t]
\begin{tabular}{@{}rllllllll@{}}
\toprule
\multicolumn{1}{c}{Step} & \multicolumn{1}{c}{T-AC (SD)} & \multicolumn{1}{c}{M-AC (SD)} &  \multicolumn{1}{c}{P (SD)} & \multicolumn{1}{c}{R (SD)} \\ \midrule
0  & 87.1 (10.7) & 97.4 (2.8) & 69.3 (12.3) & 90.2 (9.6) \\
1  & 60.0 (22.3) & 92.7 (6.6) & 45.7 (22.6) & 78.0 (19.2) \\
2  & 49.3 (22.9) & 92.4 (6.4) & 36.9 (20.8) & 68.3 (22.0) \\
3  & 40.8 (26.0) & 90.2 (8.6) & 31.4 (19.5) & 61.4 (22.3) \\
4  & 35.7 (27.5) & 87.8 (9.8) & 27.6 (19.6) & 55.8 (21.9) \\
5  & 34.3 (28.19 & 86.8 (9.2) & 26.3 (19.7) & 54.5 (23.9) \\
\bottomrule
\end{tabular}
\caption{The evaluation of the estimates when the tempo grows (or decreases) by Step milliseconds every 16th note.}
\label{tab:increment}
\vspace{-10mm}
\end{table}

\subsubsection{Discussion}
The results show that the system cannot maintain its stability even if the tempo change is performed over a period of time rather than immediately. If the change is limited, we can expect the performance to degrade in a limited way for what concerns the tempo estimate. The estimate of the next measure's onset becomes less and less effective, because the previsions that are made before the tempo change cannot account for the final tempo change. 
This is not really surprising: even for a human musician it is hard to follow a crescendo/rallentando only by hearing another musician perform it, without having practiced it before or without having a visual cue (like a director).

\vspace{-3mm}
\section{Conclusions}
\label{sec:conclusion}

In this paper, we described a method for the detection of beat duration, meter and onset of a measure in real-time, that makes minimal assumptions on the input. This system should ideally allow software to interact in real-time with a human rhythmic improvisation, without the need to limit the human player to a prefixed meter or tempo, and without the need of leaving the synchronization to the human. 

We described how we evaluated the effectiveness of the algorithm via simulation, using quantitative metrics. The algorithm can perform in real-time if the number of notes to analyze is kept under a certain threshold. The accuracy of the algorithm, on the other hand, was low in many settings. The system performs best around the tempo that is preferred by most human players (120~bpm), meaning that, despite the low results in other settings, this algorithm could still be useful in some applications even without adding more sophisticated polishing of the estimates. 

To obtain the desired outcome and be effectively able to adapt to any human improvisation without needing to limit it beforehand, further developments are needed. 
One direction for research would be to give the system a better way to tell if its estimate is precise. 
We briefly described clarity (see Algorithm~\ref{alg:beat}), that is a first metric to describe how confident the system feels in its prediction, but it was not used in the evaluation of the system since further research is needed to correctly interpret the meaning of this metric. 
Other such metrics could be added, so that an application using this system could discard uncertain estimates of tempo and meter, hopefully obtaining more stability.
Another necessary development concerns the ability of the system to adapt in real-time in order to better consider changes in tempo and meter. For example, the system could benefit from shorter time windows analysis along with the main one, to detect sudden changes that would  be overlooked by looking at a longer time period. 

This system was already used in an application that creates a musical accompaniment~\cite{carnovalini_multilayered_2019,simonetta_symbolic_2018} to a rhythmic improvisation~\cite{carnovalini}. In that case, the synchronization was perceived as good by the users, probably also because as soon as a musical output starts, the human players will tend to adapt to the tempo they hear, especially if that tempo is close to the one they were already playing. 
In general, it is possible to use the techniques here described adapting them to make use of any additional data provided by the system they are being used for, possibly leading to better performances and opening the way for many rhythm-based Human-Computer Interaction applications.